\documentclass[prb,superscriptaddress,twocolumn,preprintnumbers,a4,floatfix,longbibliography]{revtex4-1}
\usepackage{amsmath, amssymb, psfrag, tabls, dcolumn, bm, color}
\usepackage[sort&compress]{natbib}
\usepackage[pdftex]{graphicx}
\usepackage{graphicx}
\usepackage{bm}

\newcommand\smallurl[1]{{\tiny \url{#1}}}

\newcommand{\be}{\begin{equation}}
\newcommand{\ee}{\end{equation}}
\newcommand{\bea}{\begin{equation*}}
\newcommand{\eea}{\end{equation*}}
\newcommand{\ba}{\begin{array}}
\newcommand{\ea}{\end{array}}
\newcommand{\beqa}{\begin{eqnarray}}
\newcommand{\eeqa}{\end{eqnarray}}
\newcommand{\beqaa}{\begin{eqnarray*}}
\newcommand{\eeqaa}{\end{eqnarray*}}

\newcommand{\matr}{\left( \begin{array}}
\newcommand{\ematr}{\end{array} \right)}

\newcommand{\rb}{\mbox{\boldmath $r$}}
\newcommand{\Rb}{\mbox{\boldmath $R$}}

\newcommand{\vv}{\mbox{$\vec{v}$}}

\newcommand{\der}{{\rm d}}

\newcommand{\lsim}{{\;\raise0.3ex\hbox{$<$\kern-0.75em\raise-1.1ex\hbox{$\sim$}}
\;}}
\newcommand{\gsim}{{\;\raise0.3ex\hbox{$>$\kern-0.75em\raise-1.1ex\hbox{$\sim$}}
\;}}

\def\sop{\mathcal{S}}
\def\sopt{\mathcal{S}_t}

\def\dsop{\hat{D}(\mathcal{S})}
\def\dsopt{\hat{D}(\mathcal{S}_t)}
\def\dsopn{\hat{D}(\mathcal{S}^n)}
\def\dsopN{\hat{D}(\mathcal{S}^N)}
\def\mos{Mo$_6$S$_6$}
\def\epso{\epsilon_0}
\def\egap{\mathcal{E}}
\def\anharmonic{\beta}

\def\urlprefix{}
\def\url#1{}

\def\doi#1{}

\begin{document}

\title{Quantum Simulations of One-Dimensional \\Nanostructures under Arbitrary Deformations}

\author{Pekka Koskinen}
\email[email:]{pekka.koskinen@iki.fi}
\address{NanoScience Center, Department of Physics, University of Jyvaskyla, 40014 Jyvaskyla, Finland}

\begin{abstract}
A powerful technique is introduced for simulating mechanical and electromechanical properties of one-dimensional nanostructures under arbitrary combinations of bending, twisting, and stretching. The technique is based on a novel control of periodic symmetry, which eliminates artifacts due to deformation constraints and quantum finite-size effects, and allows transparent electronic structure analysis. Via density-functional tight-binding implementation, the technique demonstrates its utility by predicting novel electromechanical properties in carbon nanotubes and abrupt behavior in the structural yielding of Au$_7$ and \mos\ nanowires. The technique drives simulations markedly closer to the realistic modeling of these slender nanostructures under experimental conditions.
\end{abstract}

\maketitle

\section{Introduction}
A significant part of contemporary nanomaterial research investigates one-dimensional (1D) nanostructures. Research motivations originate from a plethora of applications among medicine\cite{baughman_science_02}, nanoelectronics\cite{ohnishi_nature_98,jonsson_NT_04,Popov2008,Rong2014}, nanomechanics\cite{baughman_science_99,lima_science_12,foroughi_science_11}, filters\cite{zhang_nanolett_2016}, sensors\cite{collins_science_00,Cheng2015}, material reinforcement\cite{thorstenson_CST_2001}, and the tailoring of material properties.\cite{wang_MSE_2008,kuc_AM_2009} Some 1D nanostructures are synthesized bottom-up, others fabricated top-down\cite{charlier_science_97,tu_nature_09,Ding2009,Jin2010,Yu2016a,hakkinen_JPCB_00}, and some are simply found directly in the nature.\cite{watson_nature_1953,reibold_nature_06} Yet all these nanostructures share one common feature: extreme slenderness. Due to large aspect ratios, they are prone to bending, twisting, and stretching, along with their arbitrary combinations. Such deformations are ubiquitous in practice, as proven by numerous experiments.\cite{hertel_PRB_98,Philp2003,Golberg2007,Xu2009,Zhu2014}



But just as eliminating deformations in experiments is hard, incorporating them into theory is even harder. All deformations are possible in finite structures, but related simulations are plagued by problems. First, most nanostructures have so many atoms that their straightforward simulation is simply out of question. Second, quantum simulations of finite structures are often deteriorated by finite size artifacts. Third, unless specifically designed to mimic experimental settings, mechanical deformation constraints can sabotage the very phenomena under study. And fourth, electronic structure analysis of finite constrained structures is often cumbersome. Therefore, 1D nanostructures are best treated by periodic boundary conditions, effectively modeling infinite extensions. Various loading conditions have been investigated earlier, but arbitrary deformations have always been simulated using finite structures.\cite{Yakobson1996,Enyashin2007} And although recent methodological advances have enabled simulating also periodic structures under pure twisting and pure bending \cite{dumitrica_JMPS_07,cai_JMPS_08,koskinen_PRL_10,kit_PRB_11},  experimental deformations are rarely pure.\cite{warner_nmat_11} Electronic structure simulations of 1D nanostructures with realistic, arbitrary deformations have remained elusive. 


In this Article, therefore, I introduce a technique to model 1D nanostructures with arbitrary deformations. Based on revised periodic boundary conditions, it eliminates artifacts related to quantum finite-size effects and mechanical deformation constraints. It also allows easy electronic structure analysis in studies on electromechanics. I demonstrate the utility of the technique by revealing surprises in the electromechanical properties of carbon nanotubes and by predicting unconventional structural yielding behavior in Au$_7$ and \mos\ nanowires.

\section{The Technique}

\subsection{1D Periodicity with Customized Symmetry}

To first introduce notations, consider electrons in a potential $V(\rb)$ invariant under an isometric symmetry operation $\dsop V(\rb)=V(\sop^{-1}\rb)=V(\rb)$, where $\rb'=\sop \rb$. Since the Hamiltonian operator $\hat{H}=-\hbar^2/(2m_e)\nabla^2+V(\rb)$ commutes with the operator $\dsop$, the two operators share the same eigenstates. If we further assume periodic boundary condition after $N$ successive operations of $\dsop$, that is $\dsopN\equiv \hat{1}$, the eigenstates acquire the property $\dsopn \psi_{a\kappa}(\rb) = \exp(i\kappa n)\psi_{a\kappa}(\rb)$, 
where $n$ is the number of symmetry operations and $\kappa \in [-\pi,\pi]$ is a good quantum number used to index the eigenstates. Consequently, the wave function $\psi_{a\kappa}(\rb)$ within a minimal unit cell determines the wave function in the entire extended structure. If the symmetry operation $\sop$ is the translation $\mathcal{T}$, the above is obviously nothing but Bloch's theorem in one dimension.\cite{bloch_ZP_29} The theorem however applies also for symmetries beyond translation, although as yet few implementations exploit this feature.\cite{mintmire_PRL_92,dumitrica_JMPS_07} The formalism of using symmetry and periodicity in this generalized fashion is referred to as revised periodic boundary conditions (RPBC).\cite{koskinen_PRL_10,kit_PRB_11}

\begin{figure}[t]
\includegraphics[width=0.8\columnwidth]{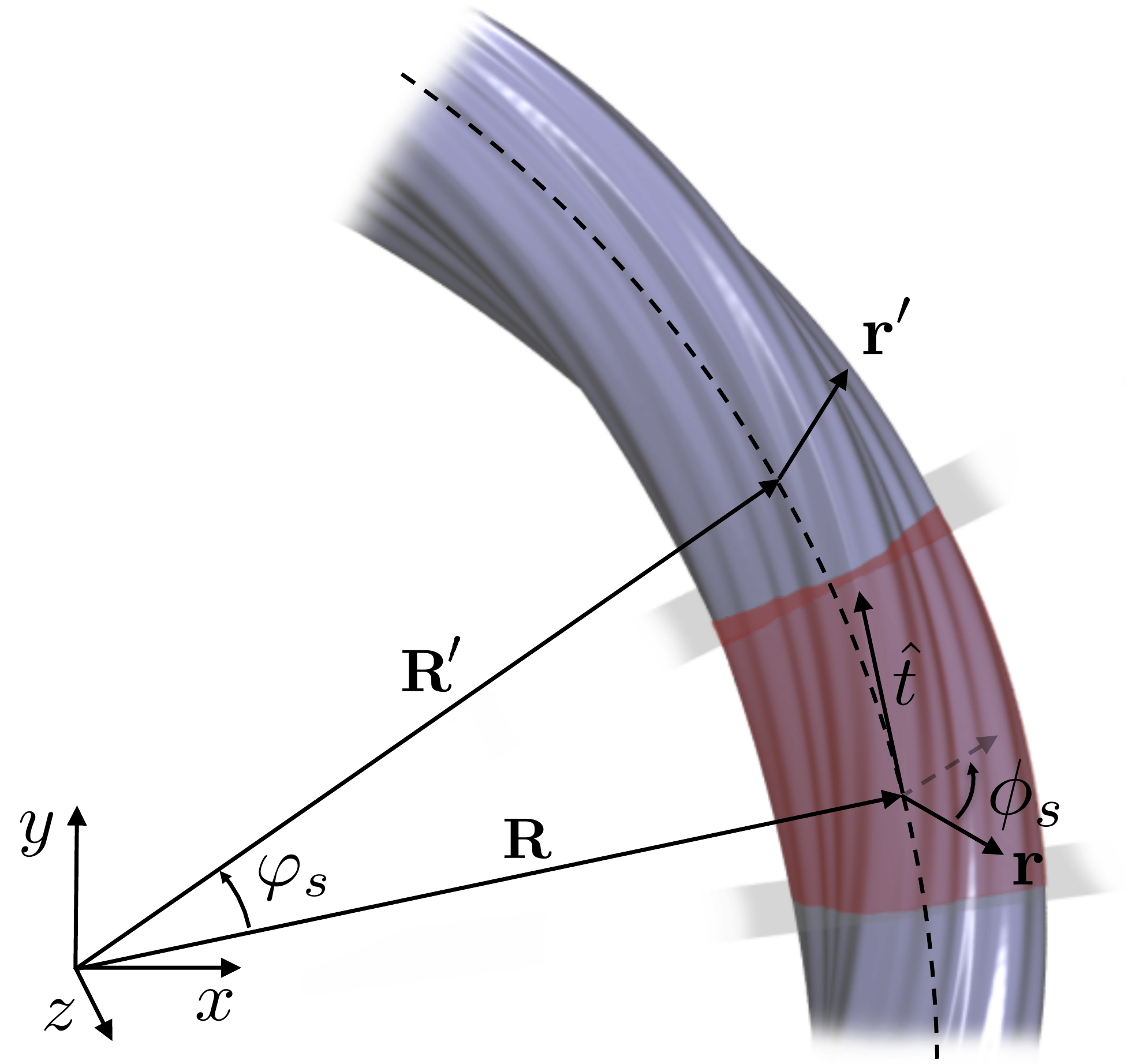}
\caption{Deformed 1D nanostructure. Atoms obey the symmetry operation $\rb' = \sopt \rb$ [Eq.~(\ref{eq:s})]. Arbitrary deformations can be created by simultaneous adjustment of bending (via $\varphi_s$ and $R=|\Rb|$), twisting (via $\phi_s$), and stretching (via $R$ and $\varphi_s$). The chiral symmetry axis (dashed curve) lies in the $xy$-plane. The red shaded area sketches the simulation cell.}
\label{fig:sketch}
\end{figure}

Consider then a one-dimensional nanostructure and the symmetry operation 
\begin{equation}
\rb' = \sopt \rb \equiv \mathcal{R}[\hat{z}(\varphi_s)](\Rb + \mathcal{R}[\hat{t}(\phi_s)](\rb-\Rb)), 
\label{eq:s}
\end{equation}
where $\mathcal{R}[\hat{u}(\theta)](\vv)$ means rotating a vector $\vv$ an angle $\theta$ with respect to the axis $\hat{u}$, where the angles $\varphi_s$ and $\phi_s$ are small (Fig.~\ref{fig:sketch}). Successive operation of $\sopt$ thus stands for a bending operation around $z$-axis combined to a screw operation around gradually reorienting chiral axis with a radius of curvature $R=|\Rb|$. Compared to previous usage of symmetry, the operation (\ref{eq:s}) conceals one fundamentally novel feature: position-dependence of the operation itself. This dependence makes the operation in principle non-isometric. It will turn out, however, that if either $\varphi_s$ or $\phi_s$ or both are small and if atom positions $\rb_I$ in the nanostructure have the symmetry $\rb'_I=\sopt \rb_I$, the property of isometry, the symmetry of the potential $V(\sopt \rb)\approx V(\rb)$, and the commutation of operators $[\dsopt,\hat{H}]=0$ become valid approximations. Due to the commutation of operators, in the spirit of revised periodic boundary conditions, $\psi_{a\kappa}(\rb)$ within a unit cell then suffices to describe the entire extended nanostructure (Fig.~\ref{fig:sketch}).\cite{koskinen_PRL_10,kit_PRB_11} Although physically sensible structures require that $\varphi_s$ and $\phi_s$ be integer fractions of $2\pi$, in practice their smallness renders such requirements irrelevant. That is, by requiring an interaction range small compared to $R$, bending, twisting, and stretching can be regarded as \emph{local} deformations, as established earlier.\cite{malola_PRB_08b,kit_PRB_12,koskinen_PRB_10b}

Consequently, Eq.(\ref{eq:s}) becomes the foundation that enables effective modeling of 1D nanostructures with arbitrary deformations. Deformations are controlled via the parameters $R$, $\varphi_s$, and $\phi_s$ as follows. First, by considering a structure with diameter $D$, bending can be quantified by the strain $\Theta=D/(2R)$. In the absence of axial strain the bending angle is $\varphi_s=L_0/R$, where $L_0$ is the cell length of the undeformed structure, and $\Theta$ equals the maximal tensile and compressive strains along the tangential direction. Second, chiral twisting can be quantified by sidewall shear $\gamma=(D/2)\phi_s/L_0$. Third, axial strain can be quantified simply by $\varepsilon=(L-L_0)/L_0$, where $L=R\varphi_s$ is the strained axial length. Thus, the three strains $(\Theta,\gamma,\varepsilon)$ fully quantify arbitrary local deformations in 1D nanostructures. Note that deformations are created not by external constraints but by underlying symmetries; all atoms remain fully unconstrained.

\subsection{Density-Functional Tight-Binding Implementation}

I implemented the technique using density-functional tight-binding (DFTB) method and the \textsc{hotbit} code\cite{porezag_PRB_95,koskinen_CMS_09}. Any classical force field or electronic structure method would have suited equally well, but the DFTB formalism allows straightforward implementation and describes the energetics and electronic structures of covalent and even metallic systems with reasonable accuracy.\cite{porezag_PRB_95,Seifert2000,koskinen_NJP_06} The pertinent parametrizations were adopted from Refs.~\onlinecite{porezag_PRB_95}, \onlinecite{Seifert2000}, and \onlinecite{koskinen_NJP_06}.

According to the RPBC formalism\cite{koskinen_PRL_10,kit_PRB_11}, the DFTB electron wave functions under the symmetry (\ref{eq:s}) are described by the revised Bloch basis functions 
\begin{equation}
|{\kappa},\mu \rangle \equiv \varphi_{\mu}({\kappa},\rb)=\frac{1}{\sqrt{N}}\sum_{n} \exp(-i{ \kappa}\cdot n)\hat{D}(\mathcal{S}_t^{n}) \varphi_\mu(\rb),
\label{eq:bloch-basis}
\end{equation}
where $\varphi_\mu(\rb)$ are minimal set of local orbitals and $\sum_{n}1=N$ is the number of unit cells. In this basis the Hamiltonian is diagonal in ${\kappa}$,
\begin{equation}
\langle {\kappa},\mu|\hat{H}|{\kappa'},\nu\rangle=\delta({\kappa}-{\kappa'})\sum_{n} \exp(-i{\kappa}\cdot n)H_{\mu\nu}({n}),
\end{equation}
where the Hamiltonian matrix elements are
\begin{equation}
H_{\mu\nu}({n})=\int \varphi_\mu^*(\rb)\hat{H} \left[ \hat{D}(\mathcal{S}_t^{n})\varphi_\nu(\rb) \right] \der^3 r.
\label{eq:h-mel}
\end{equation}
Together with analogous equations for overlap matrix elements, the DFTB total energy expression is as usual\cite{porezag_PRB_95,koskinen_CMS_09}, forces are calculated as parametric derivatives of the total energy, and structural relaxation and molecular dynamics are performed in the usual fashion.\cite{koskinen_CMS_09} The positions of atoms' periodic images were mapped exactly via Eq.~(\ref{eq:s}), but for simplicity the orbital rotations $\hat{D}(\mathcal{S}^{n})\varphi_\nu(\rb)$ in Eq.~(\ref{eq:h-mel}) were done using the averaged tangential vector of $\hat{t}=-\sin(\varphi_s/2)\hat{x}+\cos(\varphi_s/2)\hat{y}$. 

Finally, although the concept of a unit cell is familiar, conceptually intuitive, and visually helpful, here such a concept is in principle unnecessary. Unit cell is even sketched in Fig.~\ref{fig:sketch}, but in practical implementation it is nowhere to be found, because DFTB only requires atoms' relative positions, which are fully determined by Eq.~(\ref{eq:s}). Atoms do not need to remain inside certain spatial region. 

\subsection{Deformation Simulations}

Deformation simulations began with initial guesses for the positions of each atom $I$ that were determined from
\begin{equation}
\rb_I = \mathcal{R}[\hat{z}(\chi\varphi_s)](R\hat{x} + \mathcal{R}[\hat{y}(\chi\phi_s)](\tilde{x}_I\hat{x}+\tilde{z}_I\hat{k})), 
\label{eq:guess}
\end{equation}
where $\tilde{\rb}=\tilde{x}_I\hat{x}+\tilde{y}_I\hat{y}+\tilde{z}_I\hat{k}$ were atom positions in an undeformed 1D nanostructure centered around $y$-axis. The length of the structure was $L$ so that the variable $\chi=\tilde{y}_I/L$ ranged from $\chi=0$ to $\chi=1$. Simulations then continued either by structural relaxation using the FIRE optimizer\cite{bitzek_PRL_06} or by molecular dynamics simulations using Langevin thermostat with $10$~K temperature and $0.5$~ps damping time. 

Two notes are worth mentioning here. First, if $\phi_s=0$, then symmetry operation became
\begin{equation}
\rb' = \sopt \rb \equiv \mathcal{R}[\hat{z}(\varphi_s)](\rb)),
\label{eq:s_psi0}
\end{equation}
which made $\sopt$ independent of $R$. Therefore, in the general case, the radius of curvature for an untwisted, relaxed structure could not be controlled by $\sopt$; for structure with the symmetry $C_n$, the control could be regained by setting $\phi_s=2\pi/n$, thus mimicking the untwisted structure by a chiral symmetry operation. Second, the eigenproblem was faced with occasional technical difficulties when the simulation cell was small and orbitals interacted with their own periodic images; longer simulation cells removed these difficulties.

\section{Deformed carbon nanotubes}

\subsection{Validation of the Technique}

\begin{figure}[b]
\includegraphics[width=0.8\columnwidth]{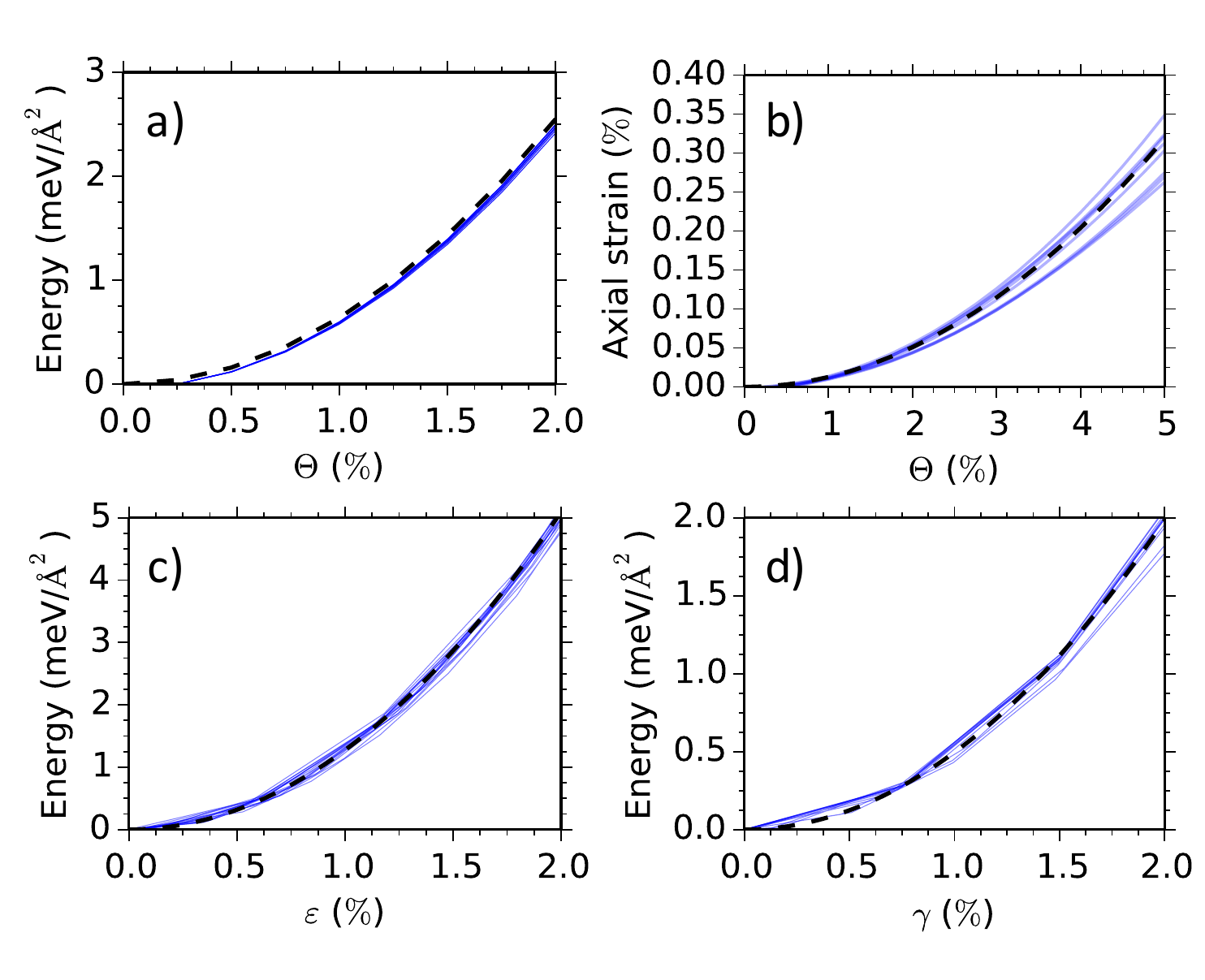}
\caption{Validity of thin sheet elasticity in carbon nanotubes under pure deformations. (The set of selected tubes are listed in Fig.~\ref{fig:systematics}). (a) Energy per sidewall area for CNTs under pure bending. The dashed line is the curve $E/A=\frac{1}{4}Y\Theta^2$. (b) Axial stretching induced by pure bending. The dashed line is the curve $\varepsilon_\Theta=\frac{3}{4}\anharmonic \Theta^2$. (c) Energy per sidewall area for CNTs under pure stretching. Dashed line is the curve $E/A=\frac{1}{2}Y\varepsilon^2$. (d) Energy per sidewall area for CNTs under pure twisting. Dashed line is the curve $E/A=\frac{1}{2}G\gamma^2$, where $G=Y/2(1+\sigma)$ is graphene's shear constant.}
\label{fig:energies}
\end{figure}

To validate the technique, I began by investigating single-walled carbon nanotubes (CNTs). They provided a good benchmark for testing, because their mechanical properties have been investigated thoroughly in the past.\cite{kudin_PRB_01,Dresselhaus2004,white_JPCB_05,Liang2006,pastewka_PRB_09,Zhao2011a} In particular, it has been shown that the energetics of carbon nanotubes with sufficiently large diameter follow closely the classical thin sheet elasticity theory with elastic parameters adopted directly from graphene.\cite{landau_lifshitz,kudin_PRB_01}

For completeness and for later reference, before proceeding with arbitrary deformations, I verified the thin sheet model with pure bending, pure stretching, and pure twisting of CNTs. The deformation energies under pure deformations agree very well with pertinent analytical estimates, where the related elastic parameters, the Young's modulus $Y=25.5$~eV/\AA\ and the Poisson ratio $\sigma=0.285$, were calculated for graphene by DFTB (Fig.~\ref{fig:energies}). In addition to linear elasticity, non-linear (bond anharmonicity) effects cause CNT stretching upon bending, because bonds at inner edge compress less than bonds at outer edge elongate.\cite{koskinen_PRB_12} Assuming a strain-dependent Young's modulus of the form $Y=(1-\anharmonic \varepsilon)Y_0$ with the anharmonicity parameter $\anharmonic=1.7$,\cite{koskinen_PRB_12} the analytically calculated axial strain upon bending becomes $\varepsilon_\Theta=\frac{3}{4}\anharmonic\Theta^2$, as confirmed by simulations (Fig.~\ref{fig:energies}b).

\begin{video}[t!]
\includegraphics[width=0.6\columnwidth]{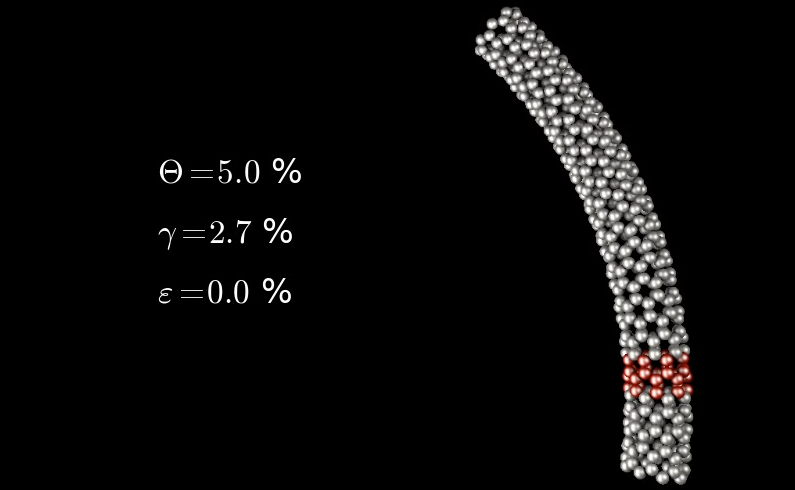}
\caption{Visualization of the deformation path in Eq.(\ref{eq:path}).}
\label{video}
\end{video}

\begin{figure}[b]
\includegraphics[width=0.6\columnwidth]{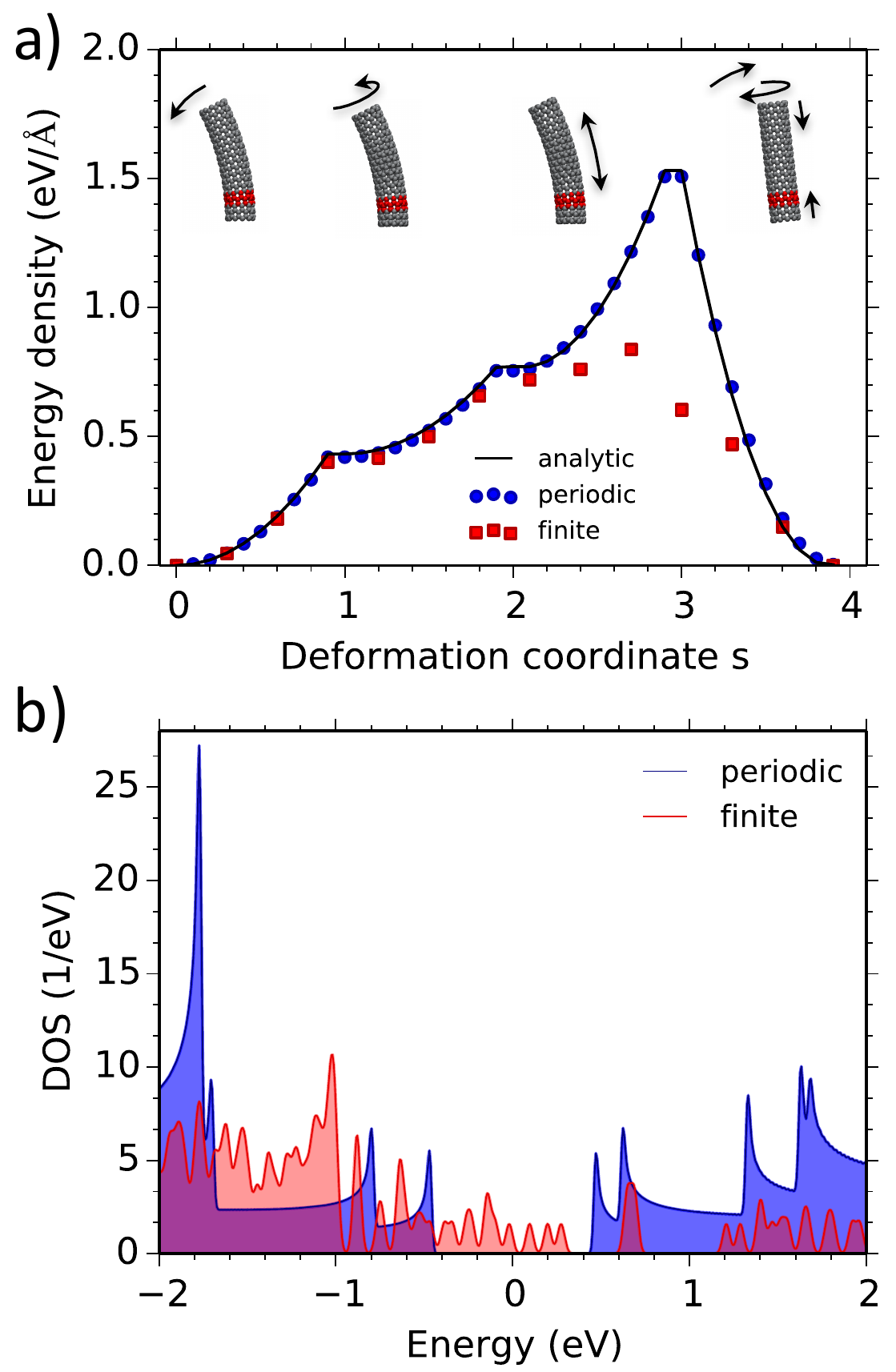}
\caption{Comparing periodic and finite calculations. (a) Energy density as a function of deformation coordinate [path of Eq.~(\ref{eq:path})] for an $(11,0)$ CNT using periodic (blue circles) and finite (red squares) simulations. Solid line corresponds to the analytic expression (\ref{eq:deformationE}); $\epso=5$~\%. Insets: periodic simulation cell consists \emph{only of the red atoms}, the gray atoms are just periodic copies; CNT of the finite simulation was $25$ times the unit cell. (b) Density of states (DOS) of undeformed $(11,0)$ CNT using periodic (blue; with $200$ $\kappa$-points) and extended but finite (red; with $1100$ atoms) simulations. Fermi-level is at zero.}
\label{fig:proof}
\end{figure}

\begin{figure*}[t!]
\includegraphics[width=\textwidth]{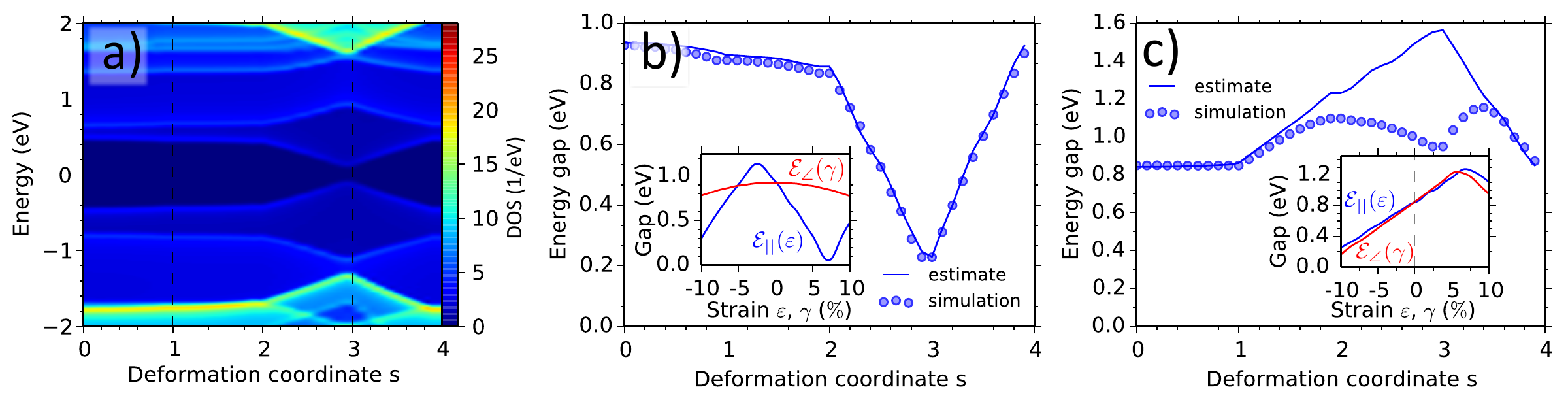}
\caption{CNT electromechanics under arbitrary deformations. (a) Contour plot for the density of states (DOS) of $(11,0)$ CNT under the deformation path of Eq.~(\ref{eq:path}) with $\epso=5$~\%. (b) Energy gap of $(11,0)$ CNT derived from panel a (circles) and from the estimate of Eq.~(\ref{eq:gap_prediction}) (solid line). Inset: energy gap of $(11,0)$ CNT as a function of pure stretching ($\egap_{||}$) and pure twisting ($\egap_\angle$). (c) Same as panel b for an $(8,4)$ CNT.}
\label{fig:electromechanics}
\end{figure*}

\vspace{0.4cm}
To create arbitrary deformations, I used a four-stage deformation path 
\begin{align}
(\Theta,\gamma,\varepsilon)(s)=
\begin{cases}
(\epso s, 0, 0) & \text{ for } s\in[0,1],\\
(\epso, \epso(s-1), 0) & \text{ for } s\in[1,2],\\
(\epso, \epso, \epso(s-2)) & \text{ for } s\in[2,3],\\
(\epso, \epso, \epso)(4-s) & \text{ for } s\in[3,4],\\
\end{cases}
\label{eq:path}
\end{align}
where $\epso$ was a maximum strain and $s \in [0,4]$ was a deformation coordinate. That is, deforming began by pure bending, proceeded by additional twisting, further by yet additional stretching, and terminated by the synchronous reversal of bending, twisting, and stretching (Video 1). The energy per unit length in a CNT under the deformation $(\Theta,\gamma,\varepsilon)$ is
\begin{equation}
E(\Theta,\gamma,\varepsilon)/L_0 = \frac{\pi D Y}{4}\left(\Theta^2+\frac{1}{1+\sigma}\gamma^2+2\varepsilon^2\right),
\label{eq:deformationE}
\end{equation}
which is a superposition of energies under separate pure deformations (Fig.~\ref{fig:energies}). Now, as demonstrated for an $(11,0)$ CNT, energy from the periodic technique agrees with Eq.~(\ref{eq:deformationE}) to high accuracy (Fig.~\ref{fig:proof}a). That is, deforming CNT merely by adjusting the symmetry operation $\sopt$ gives accurate deformation energies. I emphasize that this agreement is not trivial, because in periodic quantum simulations the elastic properties are \emph{a priori} indeterminate; they emanate automatically from the electronic structure, from wave functions, and from wave function symmetries. The agreement can be therefore considered as a direct validation for the technique and a verification of the underlying approximations.



\subsection{Comparison with Finite CNTs}

For comparison, I applied the deformation path (\ref{eq:path}) also to a finite $(11,0)$ CNT containing $25$ unit cells and $1100$ atoms. The deformation was constrained by fixing atoms $I_\text{bot}$ near bottom end and constraining the atoms $I_\text{top}$ near top end to move along a trajectory $\rb_{I_\text{top}} = \sopt^{24} \rb_{I_\text{bot}}$. For the finite CNT, Eq.~(\ref{eq:deformationE}) describes the energy well at small $s$ but poorly at large $s$. Around $s\approx 3$ the comparison of energies becomes even questionable, because the end constraints of the finite tube could not retain the deformation homogeneous and the tube axis lost its circular arc form (Fig.~\ref{fig:proof}a). In addition, a single energy and force evaluation step took some thousand times longer for the finite CNT ($\Delta t\sim 0.5$~h) than for the periodic CNT ($\Delta t\sim 2$~s), and also the number of optimization steps for finite CNT was about ten times larger.

However, the main difference and the true power of the technique lies in the electronic structure analysis. Although the atom count in the finite $(11,0)$ CNT was $1100$, the resulting aspect ratio of $12.4$ was petty compared to the experimental ratios $10^{2-4}$ or even $10^8$.\cite{zhang_ACSNano_2013} As a result, the density of states (DOS) from finite simulation was unreliable, as it included spurious end-localized states that arose from quantum finite-size effects (near the Fermi-level in Fig.~\ref{fig:proof}b). On the contrary, the electronic structure of periodic CNT could always be converged by a sufficient number of $\kappa$-points.





\subsection{Electromechanics under Arbitrary Deformations}

The DOS for the $(11,0)$ CNT (Fig.~\ref{fig:proof}b) was recorded also for the entire deformation path. Upon deforming the van Hove singularities remain prominent, but shift to higher and lower energies (Fig.~\ref{fig:electromechanics}a). These shifts are reflected in changes of the fundamental energy gap, as reported earlier for pure deformations.\cite{kane_PRL_97,yang_PRL_00,koskinen_PRB_10,koskinen_APL_11,koskinen_PRB_12} Regarding arbitrary deformations, it turned out that the gap is well described by
\begin{equation}
\egap(\Theta,\gamma,\varepsilon)=\egap(0,0,0) + \Delta \egap_{||}(\varepsilon_\Theta + \varepsilon)+\Delta \egap_{\angle}(\gamma),
\label{eq:gap_prediction}
\end{equation} 
where $\Delta \egap_{||}(\varepsilon)=\egap_{||}(\varepsilon)-\egap_{||}(0)$
is the gap change under pure axial strain and $\Delta \egap_\angle(\gamma)$ corresponding gap change under pure twist (Fig.~\ref{fig:electromechanics}b). Eq.~(\ref{eq:gap_prediction}) includes also contribution from axial strain due to bond anharmonicity, as discussed above.\cite{koskinen_APL_11,koskinen_PRB_12} This effect is visible at $s<1$, where $\gamma=\varepsilon=0$ but $\epsilon_\Theta\neq 0$ (Fig.~\ref{fig:electromechanics}b). Thus, gap changes under arbitrary deformations are given by linear superposition of gap changes in separate \emph{pure} deformations. Because the superposition is valid for all van Hove singularities, Eq.~(\ref{eq:gap_prediction}) is expected to pertain to optical transitions as well.\cite{koskinen_PRB_10} The validity of linear superposition could be anticipated, but it has never been demonstrated directly.

\begin{figure*}[t!]
\includegraphics[width=\textwidth]{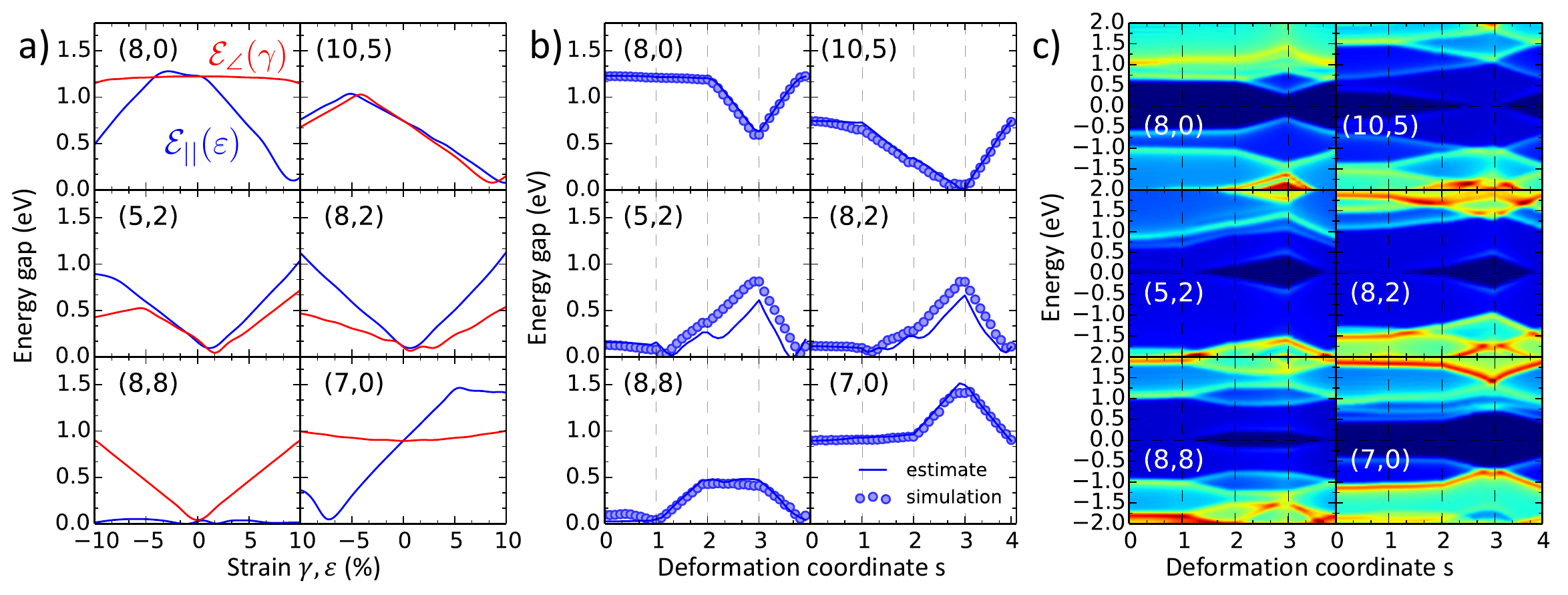}
\caption{Electromechanics of various CNTs under arbitrary deformations. a) Energy gap under pure shear (red lines) and pure stretch (blue lines), defining the functions $\mathcal{E}_{||}(\varepsilon)$ and $\mathcal{E}_{\angle}(\gamma)$. (b) Energy gaps under the deformation path of Eq.~(\ref{eq:path}). Solid lines are the estimates from Eq.~(\ref{eq:gap_prediction}), using the data of panel a. (c) Contour plot for the density of states under the deformation path of Eq.~(\ref{eq:gap_prediction}). Color scale is the same as in Fig.~\ref{fig:electromechanics}a. The parameter $\epso=5$~\% in panels b and c. }
\label{fig:systematics}
\end{figure*}

The news is, however, that sometimes superposition principle goes awry without a warning. When considering $(8,4)$ CNT under the path (\ref{eq:path}) with $\epso=5$~\%, after $s\gsim 2$ the gap starts to behave \emph{opposite} as compared to Eq.(\ref{eq:gap_prediction}) (Fig.2c). In $(8,4)$ CNT pure stretching increases the gap (inset of Fig.2c), but the stretching of already bent-and-twisted tube \emph{decreases} the gap. Combination of deformations creates synergy that causes non-linear response in the electronic structure and invalidates linear superposition. For $(8,4)$ CNT the validity of Eq.~(\ref{eq:gap_prediction}) was regained by decreasing $\epso$ down to $2$~\%\ (not shown), but the limits of validity could not be anticipated from CNT electromechanics under separate pure deformations; confirmation of possible validity required explicit simulations with arbitrary deformations. 

I repeated the above analysis with $\epso=5$~\%\ for a set of different chiral and non-chiral CNTs. It turned out that the superposition Eq.~(\ref{eq:gap_prediction}) is most accurate for zigzag and armchair tubes, and slightly less accurate for chiral tubes (Fig.~\ref{fig:systematics}). The origin for this behavior is unknown and requires further investigation.

\subsection{Poynting Effect}

\begin{figure}[b]
\includegraphics[width=0.7\columnwidth]{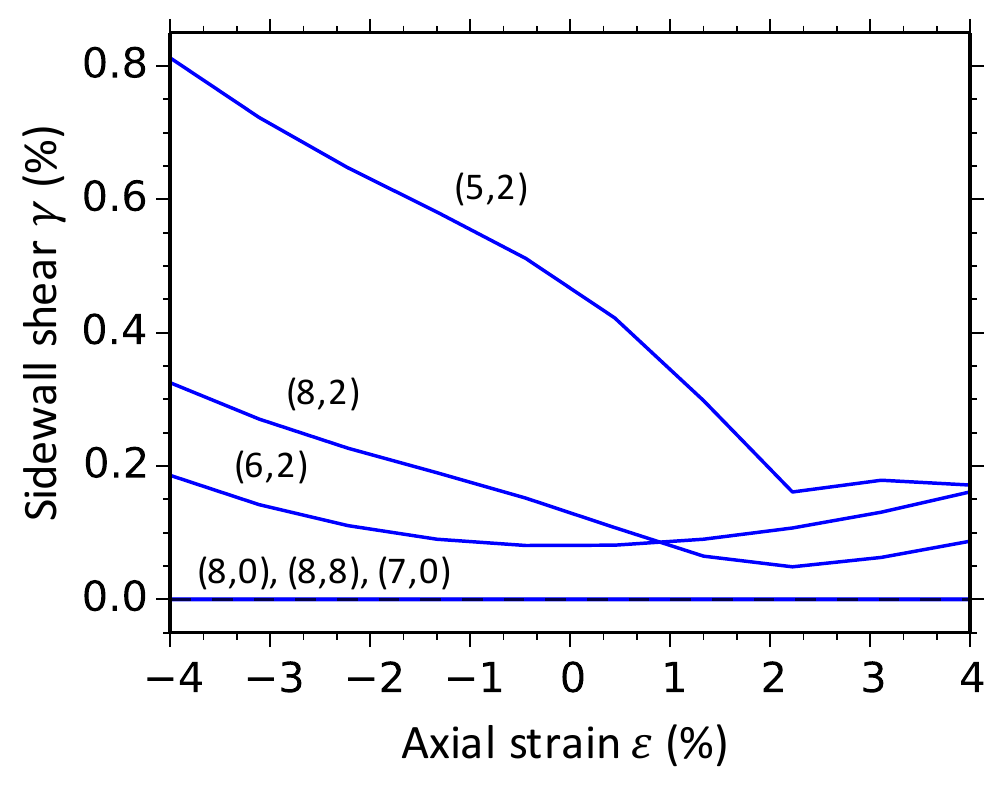}
\caption{Poynting effect in CNTs. Panels show the sidewall shear as a function of axial strain. The effect is observed only in chiral tubes, not in non-chiral zigzag ($n,0$) or armchair ($n,n$) tubes.}
\label{fig:poynting}
\end{figure}

Returning to the mechanical properties of CNTs, previous simulations have already shown that chiral tubes display the Poynting effect, which means that stretching and twisting are coupled (twisting induces stretching or the other way around).\cite{Liang2006} While predictions have been made using purely classical models, here the predictions were confirmed by a quantum-mechanical method (Fig.~\ref{fig:poynting}).

Given the Poynting effect, the bending-induced axial stretching (Fig.~\ref{fig:energies}b), and the technique to simulate arbitrary deformations, it was pertinent to investigate whether also bending could induce twisting in CNTs. The twisting angle due to bending can be roughly estimated as $\Delta \gamma = |\partial \gamma/\partial \varepsilon| \times \frac{3}{4}\anharmonic \Theta^2$, where $|\partial \gamma/\partial \varepsilon|$ measures the magnitude of the Poynting effect. At maximum the magnitude is $|\partial \gamma/\partial \varepsilon|_{\text{max}} \approx 0.1 $ (Fig.~\ref{fig:poynting}), so the largest amount of bending-induced twisting becomes $\Delta \gamma \approx 0.13 \times \Theta^2$, which turns out to be only $\Delta \gamma \sim 0.1$~\%\ even at bending as large as $\Theta=10$~\%. A CNT with $D=1$~nm bent to $\Theta=1$~\%\ would then need to be at least a quarter of a millimeter long to twist a full turn. Being this tiny in magnitude, bending-induced twisting could not be resolved in periodic simulations. Zhao and Luo\cite{Zhao2011a} reported twisting-induced bending, but it was due to the Poynting effect combined to a constrained length for a finite tube; \emph{intrinsic} bending-induced stretching in CNTs seems to be too minor to be of practical significance.

\section{Deformed M\lowercase{o}S and A\lowercase{u} nanowires}

\subsection{Electromechanics of Mo$_6$S$_6$ nanowire}

\begin{figure}[tb]
\includegraphics[width=0.85\columnwidth]{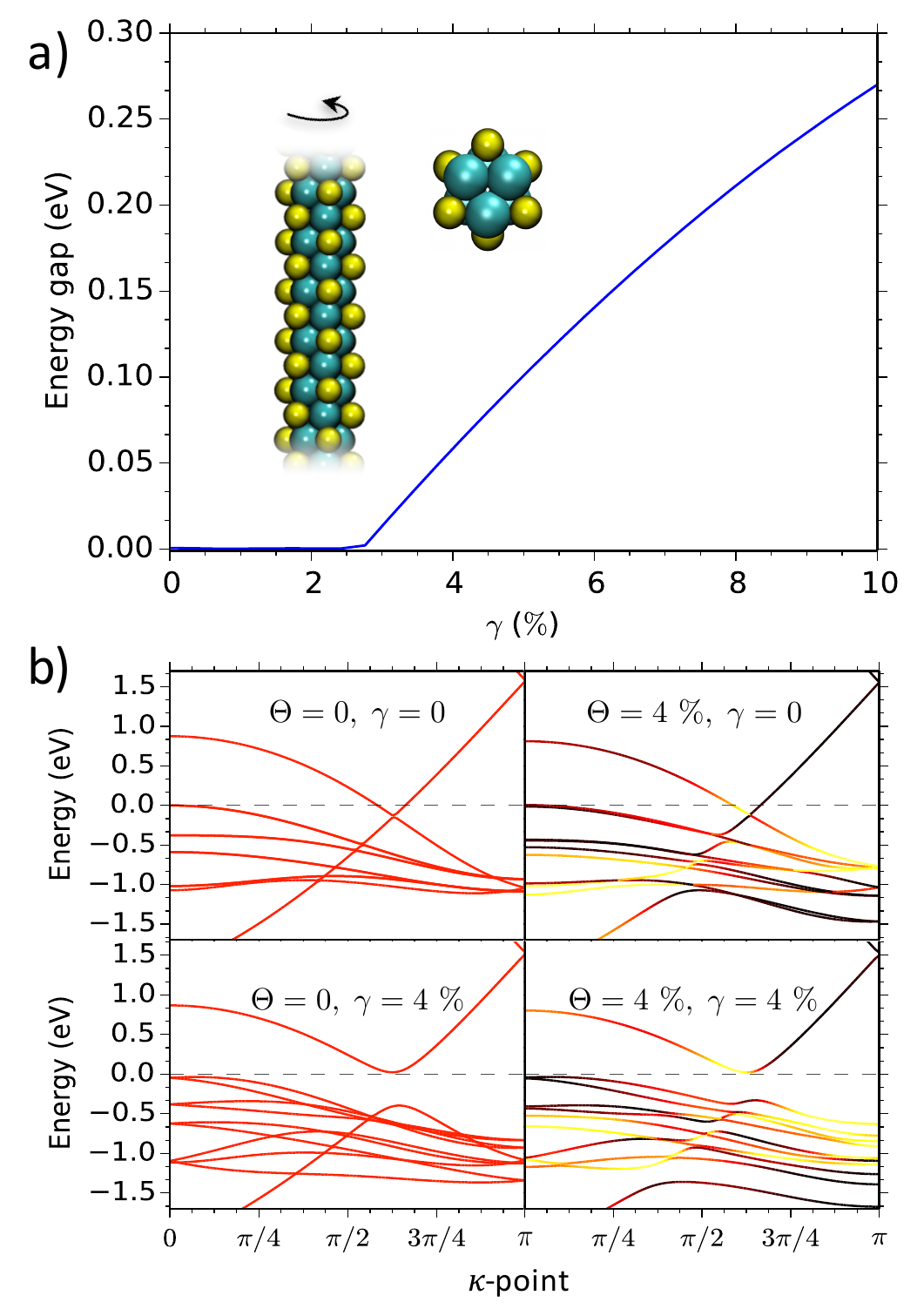}
\caption{Electromechanics of \mos\ nanowire. a) Energy gap under pure twist (with $D=0.3$~nm). Inset shows the side and cross-section views of the structure. b) Band structure under various deformations. Panels correspond to straight (upper left), purely bent (upper right), purely twisted (lower left), and bent and twisted (lower right) wires. For bent wires the coloring corresponds to wave functions localized more toward inner (darker) and outer (brighter) sides of the wire. Dashed line is the Fermi level.}
\label{fig:mos}
\end{figure}

Let us next leave CNTs aside and move to studying MoS$_2$ monolayer -derived \mos\ nanowires.\cite{Seifert2000,ghorbani-asl-SC_2013,Leen2015,Lin2016} These wires are a timely example of deformed 1D nanostructures, as demonstrated by aberration-corrected transmission electron micrographs.\cite{Lin2016} For example, the work of Lin \emph{et al.} showed highly resilient \mos\ nanowires bent up to $\Theta\sim 6-10$~\%.\cite{Lin2014a} Undeformed \mos\ wires are metallic, but twist of magnitude $\gamma=2.7$~\%\ (assuming wire diameter $D=0.3$~nm) has been predicted to open a gap, which was confirmed also here (Fig.~\ref{fig:mos}a).\cite{Popov2008} As suggested before, these properties could be exploited in an electromechanical switch that allows current propagation in a straight wire but not in a twisted one.\cite{Popov2008} Here simulations agree with the previous results under pure bending and pure twisting, but the analysis is more transparent as the band structure can be always plotted for the same minimal cell. Analysis reveals that pure bending creates small energy splittings due to weak wave function localization at inner and outer sides of the wire (Fig.~\ref{fig:mos}b). Analogous localization has been reported also in the vibrational modes of bent CNTs.\cite{malola_PRB_08b} Pure bending affects band structure weakly, but a pre-existing twist enhances the effect of bending notably (juxtapose the changes in Fig.~\ref{fig:mos}b from upper left to upper right with the changes from lower left to lower right). However, although the twisting-induced metallic-to-semiconducting transition is here seen also under bending, robust electromechanical switching operation should require also structural robustness; this is what we discuss next.


\subsection{Mechanical stability of Mo$_6$S$_6$ nanowire}

\begin{figure}[t!]
\includegraphics[width=0.8\columnwidth]{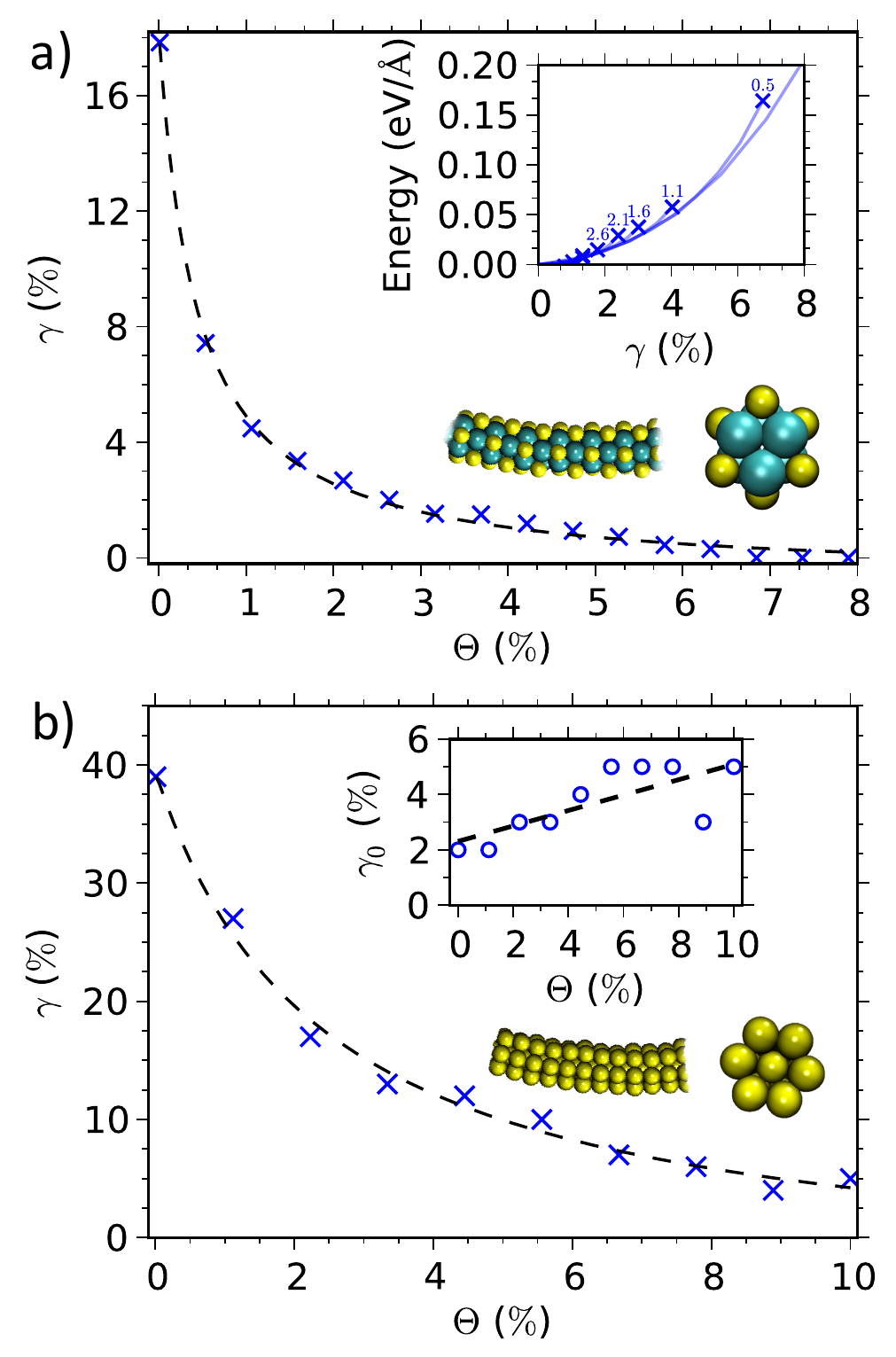}
\caption{Yield limits in \mos\ and Au nanowires under combined bending and twisting. (a) Twisting yield limit as a function of bending in the \mos\ nanowire. Dashed line is a fit discussed in the main text. Inset: Elastic energy as a function of twist for different bends $\Theta$ (numbers shown); yield points are marked by crosses. (b) Twisting yield limit as a function of bending in Au$_7$ nanowire. Inset: Minimum-energy twists for bent Au$_7$ wires.}
\label{fig:stability}
\end{figure}

Reliability of structural stability analysis calls for simulation cells larger than the minimal ones. With the minimal $12$-atom cell \mos\ was stable at least beyond $\Theta=\gamma>4$~\%, as shown above (lower right panel in Fig.~\ref{fig:mos}b). When the simulation cell length was extended to $2.6$~nm, however, atomic structure revealed its sensitivity to combined deformations. When twisted, the elastic energy first depended quadratically on $\gamma$ with the torsion constant $150$~eV\AA, in fair agreement with the literature.\cite{Leen2015} Upon further twisting the energy started to deflect from this quadratic trend and the wire began to yield (inset of Fig.~\ref{fig:stability}a). Most important, the deflection and yielding occurred at rapidly decreasing twist when bending increased. By plotting the yield points of twisting for different $\Theta$ it became evident that combining bending and twisting affects stability limits dramatically (Fig.~\ref{fig:stability}a). For example, the yield limit in purely twisted wire was $\gamma=18$~\%, but modest $\Theta\sim 0.5$~\%\ bending cut this limit to less than half. The metal-insulator transition and robust operation of electromechanical switch device is thus feasible only for relatively straight wires ($\Theta<1.5$~\%). Besides, temperatures higher than the one used here ($10$~K) would probably lower the yield limits even further.


Although the lowering of twisting yield limit under bending was anticipated, its abruptness was not (Fig.~\ref{fig:stability}a). The yield limit follows the ad hoc form $(\gamma+c_1)(\Theta+c_2)=c_3$, where $c_1=7.8\times 10^{-3}$, $c_2=4.2\times 10^{-3}$, and $c_3=8.1\times 10^{-4}$. This form differs radically from standard yield criteria valid for macroscopic solids, such as the von Mises criterion.\cite{VonMises} The von Mises criterion is based on fixed allowed energy density and suggests instead a form $b_1\gamma^2+b_2\Theta^2=b_3$ with constant $b_i$'s. Here the total energy and thereby energy density at yield point was not fixed, but dropped rapidly when $\Theta$ increased (inset of Fig.~\ref{fig:stability}a).  

\subsection{Mechanical stability of Au$_7$ nanowire}

I performed similar stability analysis also for a $0.6$~nm-diameter Au$_7$ nanowire with $1.7$~nm long cells.\cite{koskinen_NJP_06} This nanowire showed yield limits qualitatively similar to those of \mos\ (Fig.~\ref{fig:stability}b). Atom trajectories revealed that yielding occurred for the entire wire cross section at once, collectively, which helps to appreciate the qualitatively different behavior compared to macroscopic rods and wires. Earlier studies of \mos\ and Au wires have revealed several dislocations; these could be indirect indications for the low yield limits under combined deformations.\cite{Tian2014,Leen2015,Yu2016a} Yet this yielding behavior remains a puzzle that deserves further investigations.

As a final observation, bending and twisting in Au$_7$ turned out to be coupled (inset of Fig.~\ref{fig:stability}b). At given $\Theta$ the energy was minimized at varying twist $\gamma_0$, following $\gamma_0(\Theta)\approx 0.02+0.28\times\Theta$. That is, unlike in CNTs, in Au$_7$ bending induces twisting and twisting induces bending, as familiar from mechanical springs.\cite{Klinkel2003}


\section{Conclusions}

To conclude, I hope to have demonstrated that for a faithful modeling of the mechanical and electromechanical properties of 1D nanostructures, simple modeling with separate pure deformations is insufficient; explicit simulations of combined deformations are mandatory. Demand for such simulations grows as the control over 1D nanostructures improves. Perspective for this demand can be obtained by considering the list of related nanostructure examples, which include metal, semiconductor and molecular nanowires, DNA, polymers, single- and multiwalled CNTs, CNT ropes and bundles, nanoribbons of graphene and other 2D materials, among many others. Particularly relevant are their surface functionalizations, which are bound to cause complex deformations. 




\section{Acknowledgements} 
I thank Jyri Lahtinen for contributions at the early stages of the project, Petri Luosma for comments, the Academy of Finland for funding (Projects No. 283103 \& 251216), and CSC - IT Center for Science in Finland for computer resources.


%

\end{document}